# Caloric Determination of the Upper Critical Fields and Anisotropy of $NdFeAsO_{1-x}F_x$ Single Crystals

U. Welp, R. Xie, A. E. Koshelev and W. K. Kwok

Materials Science Division, Argonne National Laboratory, Argonne, Il 60439

P. Cheng, L. Fang and H.-H. Wen

Institute of Physics, Chinese Academy of Sciences, Beijing 100190, China

We present heat capacity measurements of the upper critical fields of single-crystal $NdFeAsO_{1-x}F_x$. In zero-magnetic field a clear step in the heat capacity is observed at $T_c \approx 47K$. In fields applied perpendicular to the FeAs-layers the step broadens significantly whereas for the in-plane orientation the field effects are small. This behavior is reminiscent of the $CuO_2$-high-$T_c$ superconductors and is a manifestation of pronounced fluctuation effects. Using an entropy conserving construction we determine the transition temperatures in applied fields and the upper critical field slopes of $\partial H_{c2}^c/\partial T = -0.72$ T/K and $\partial H_{c2}^{ab}/\partial T = -3.1$ T/K. Zero-temperature coherence lengths of $\xi_{ab} \approx 3.7 nm$ and $\xi_c \approx 0.9 nm$ and a modest superconducting anisotropy of γ ~ 4 can be deduced in a single-band model.

The recent discovery of superconductivity in LaFeAsO$_{1-x}$F$_x$ [1] has led to the emergence of a new family of layered high-temperature superconductors with compositions REFeAsO$_{1-x}$F$_x$ with rare earths RE = Sm, Ce, Nd, Pr, Gd, Tb, Dy. Currently, the highest value of T$_c$ is 55 K [2]. These materials have a layered tetragonal crystal structure [1] in which the FeAs-layers are believed to carry superconductivity whereas the REO$_{1-x}$F$_x$ layers serve as charge reservoirs. The undoped parent compounds (LaFeAsO [3], NdFeAsO [4] and CeFeAsO [5]) are semimetals that undergo a tetragonal-to-orthorhombic transition near 150 K and enter an antiferromagnetically ordered state at lower temperature. Upon electron-doping through substitution of O with F or hole-doping through substitution of RE with Sr [6] the structural and the antiferromagnetic transitions are suppressed and a superconducting state emerges resulting in an intriguing interplay of magnetism and superconductivity reminiscent of the behavior of the cuprate high-T$_c$ superconductors. Furthermore, electronic band structure calculations [7, 8] indicate that the Fermi surface contains multiple sheets derived from the Fe d-orbitals, which could give rise to multi-band superconductivity [9-11].

The upper critical field, H$_{c2}$, and its anisotropy are fundamental characteristics that shed light on these microscopic properties of the new FeAs-superconductors. To date, H$_{c2}$ has been inferred mostly from transport measurements on polycrystalline samples. Here we present the first thermodynamic determination of the upper critical fields of single-crystal NdFeAsO$_{1-x}$F$_x$ using heat capacity measurements. In zero-magnetic field a clear step in the heat capacity is observed at $T_c \approx 47K$. In fields applied perpendicular to the FeAs-layers the step broadens significantly whereas for the in-plane orientation the field effects are small. This behavior is reminiscent of the CuO$_2$-high-T$_c$ superconductors and is a manifestation of pronounced fluctuation effects. Using an entropy conserving construction we determine the transition temperatures in applied fields and the upper critical field slopes of $\partial H_{c2}^{c}/\partial T = -0.72$ T/K and $\partial H_{c2}^{ab}/\partial T = -3.1$ T/K. These values are significantly lower than previous resistive determinations [12, 13]. Zero-temperature coherence lengths of $\xi_{ab} \approx 3.7 nm$ and $\xi_c \approx 0.9 nm$ and a modest superconducting anisotropy of $\gamma \sim 4$ can be deduced in a single-band model

The crystals used in this study have a nominal composition of x = 0.18 and were grown from a NaCl-flux as described in [12]. We performed the caloric measurements using a membrane-based steady-state ac-micro-calorimeter [14]. It utilizes a thermocouple composed of Au-1.7%Co and Cu films deposited onto a 150 nm thick $Si_2N_4$-membrane as thermometer. Ten $NdFeAsO_{1-x}F_x$ crystallites of various sizes were mounted onto the thermocouple using minute amounts of Apiezon N grease (see inset of Fig. 3). An ac-heater current at 47 Hz is adjusted such as to induce oscillations of the sample temperature of 50 to 200 mK. The ac-technique is very sensitive for detecting changes in the specific heat, but less so for the determination of absolute values. This is particularly true in the present case where the signal due to superconductivity amounts to only a fraction of $10^{-3}$ of the total signal. Therefore, we concentrate here on tracing the superconducting phase boundaries and the anisotropy.

Figure 1 shows the superconducting part of the heat capacity signal for various fields applied along the c-axis and the ab-plane, respectively. These data are obtained by subtracting from each measurement the normal state background, $c_n$, which includes contributions from phonons, normal electrons and possibly magnetism, and the addenda. Here, we use the data in a magnetic field of 7.5 T applied along the c-axis for which there are no features discernable as background. This choice of background is suitable for the high temperature part of the phase diagram. In zero-field a clear step with onset near 48 K and a peak near 46 K is observed. These temperature values are in reasonable agreement with the onset and zero-resistance temperatures seen in the resistive transitions of similar crystals [12]. The absence of any additional structure in the heat capacity transition indicates that the transition width observed in our measurements is not caused by an average over several crystallites with individual sharp transitions but that it is an inherent property of the current material. With increasing field along the c-axis the heat capacity anomaly broadens significantly and its height decreases rapidly. In fields higher than 3 T || c the step has essentially disappeared and the peak has transformed into a kink that can be traced up to 6 T. In contrast, for fields applied along the ab-directions the specific heat step stays well defined and shifts slightly to lower temperatures as indicated by the dashed lines. This general behavior is similar to observations on $CuO_2$ high-$T_c$ superconductors [15], but is markedly different from $MgB_2$ [16], which has similar values of $T_c$ and anisotropy as $NdFeAsO_{1-x}F_x$.

The systematic vertical displacement of the curves in Fig. 1 is a measured effect, which indicates that the chosen background signal contains a magnetic contribution. Previous magnetization studies [17] have shown that $NdFeAsO_{1-x}F_x$ has a large paramagnetic moment that originates from the large atomic magnetic moments of $Nd^{3+}$ - the free ion moment is 3.6 $\mu_B$. The specific heat of a paramagnet decreases with field for $\mu_B B < k_B T$. Therefore, for the temperature range near $T_c$ the normal state background in zero-field is larger than in 7.5 T resulting in the positive off-sets.

The data in Fig. 1 allow the construction of the anisotropic superconducting phase diagram of NdFeAsO$_{1-x}$F$_x$ as shown in Fig. 2. We define the transition temperature T$_{c2}$(H) through an entropy conserving construction as illustrated in Fig. 3. Figure 2 also shows the field dependence of the temperature T$_p$ of the peak / kink in c(T). The average slopes of H$_{c2}$ for the c-axis and ab-axes, respectively, are $\partial H_{c2}^c/\partial T$ = -0.72 T/K and $\partial H_{c2}^{ab}/\partial T$ = -3.1 T/K resulting in a modest superconducting anisotropy of γ ~ 4.3. Using the Ginzburg-Landau relations for the upper critical field, $H_{c2}^c = \phi_0/2\pi\xi_{ab}^2$ and $H_{c2}^{ab} = \phi_0/2\pi\xi_{ab}\xi_c$, and the single-band WHH expression relating the zero-temperature upper critical field to the slope at T$_c$, $H_{c2}(0) = -0.69\ T_c\ (\partial H_{c2}/\partial T)_{T_c}$, we obtain the following parameters for our NdFeAsO$_{1-x}$F$_x$ samples: $H_{c2}^c(0) \approx 23\ T$, $H_{c2}^{ab}(0) \approx 100\ T$, $\xi_c(0) \approx 0.9\ nm$, $\xi_{ab} \approx 3.7\ nm$. As a reference point, the paramagnetic limiting field H$_P$ has a value of H$_P$[T] = 1.84 T$_c$[K] = 86.5 T which is slightly below our estimate of H$_{c2}$$^{ab}$. Previous determinations of the upper critical field of NdFeAsO$_{1-x}$F$_x$ using magneto-transport measurements on polycrystalline [13] and single crystal [12] samples have yielded significantly larger H$_{c2}$-values. We attribute this difference to the uncertainty in assigning a resistivity criterion for the determination of H$_{c2}$ [18] when the transitions broaden strongly in a magnetic field. We note though that the zero-resistance points for H || c obtained on similar crystals [12] coincide well with T$_p$, that is, the completion of the transition. Furthermore, the anisotropy parameter deduced from the single-crystal transport measurements [11] is close to the value obtained here. This is consistent with the observation that the anisotropy factor deduced from our caloric measurements does not depend on the definition of the superconducting transition, i.e., T$_p$ or T$_{c2}$, as demonstrated directly in Fig. 3. The data for the entire transition taken in 0.5 T ∥ c superimpose very well onto those taken in 2 T ∥ ab, indicating that the anisotropy is close to γ ~ 4. Our data do not reveal – at least in the temperature range close to $T_c$ presented here – a temperature dependence of γ as has been recently observed in torque measurements on $SmFeAsO_{1-x}F_x$ [11] and in the flux flow resistance of $NdFeAsO_{1-x}F_x$ [18]. Our findings of a temperature independent ansisotropy of γ ~ 4 are in good agreement with results of recent rf-screening experiments [20].

The angular dependent transition temperature $T_{c2}(\theta)$ is given by $T_{c2}(\theta) = T_{c0} + H\sqrt{\cos^2(\theta) + \gamma^2 \sin^2(\theta)} \Big/ \left(\partial H_{c2}^{ab}/\partial T\right)$ within the effective mass model of the Ginzburg-Landau theory of anisotropic superconductors, assuming linear phase boundaries. Here, θ is the angle of the magnetic field with respect to the FeAs-planes and $T_{c0}$ is the zero-field transition temperature. Fig. 4 shows the angular dependence of the transition temperature in 1.5 T together with a fit according to the effective mass model. Within the experimental uncertainties the data are well described by an anisotropy coefficient of γ ~ 4, consistent with the discussion above. A mosaic spread due to the mounting of the crystallites could cause a reduced measured anisotropy. However, we estimate the possible misalignment to be less than 5 deg, which would not affect the determination of γ in a significant way.

Electronic band structure calculations [7] have revealed five Fermi surface sheets, 2 high-velocity cylindrical electron sheets centered on the MA line of the tetragonal Brillouin zone, 2 low-velocity cylindrical hole sheets centered on the ΓZ-line, and a three-dimensional heavy hole pocket centered on the Z-point. An over-all effective mass anisotropy $m_c/m_{ab} = (\xi_{ab}/\xi_c)^2 = \gamma^2$ of ~15 is found, which depends on the electron-doping level. This prediction is in good agreement with our experimental value for the superconducting anisotropy.

The evolution of the heat capacity in applied fields shown in Fig. 1 can be attributed to the effects of superconducting fluctuations. The importance of fluctuations in zero magnetic field is quantified by the Ginzburg number $G_i = \left(k_B T_{c0}/H_c^2 \xi_{ab}^2 \xi_c\right)^2/2$ = $\left(8\pi^2 k_B T_{c0} \lambda_{ab}^2 / \phi_0^2 \xi_c\right)^2/2$. $k_B$, $\phi_0$, $H_c$, and $\lambda_{ab}$ are the Boltzmann constant, flux quantum, thermodynamic critical field and in-plane magnetic penetration depth, respectively. Using the above estimate for $\xi_c$ and a typical value of $\lambda_{ab} \sim 200$ nm [20, 21] we obtain $G_i \sim 2 \cdot 10^{-3}$. This value is lower than that for $CuO_2$-high-$T_c$ superconductors ($G_i \sim 10^{-2} - 1$) but is about ten times larger than the value for single-crystal $MgB_2$. In high applied fields the transition progressively broadens due to enhanced fluctuations as expressed by a field dependent Ginzburg number $G_i(H) = \left(4\pi k_B T_{c0} H / \phi_0 \xi_c H_c^2\right)^{2/3}$ [22] which would yield a temperature range of ~2 K of strong fluctuations in a field of 5 T.

The analysis presented above, in particular the extrapolation of the zero-temperature values, is valid for a single-band BCS-superconductor. The deviations of the resistively determined upper critical field of polycrystalline $LaFeAsO_{1-x}F_x$ [9] at high fields from the expected single-band WHH variation, the temperature dependent superconducting anisotropy of $SmFeAsO_{1-x}F_x$ crystals [11] from magnetic torque and from flux flow in $NdFeAsO_{1-x}F_x$ crystals [19], as well as the presence of multiple iron d-bands at the Fermi-energy [7, 8] have lead to the suggestion that the FeAs-superconductors could be two (multiple) band superconductors. Even though the temperature dependence of $T_p$ for H ∥ c gives some indication for upward curvature, the limited temperature range and the experimental uncertainties of the present data preclude a definite conclusion.

In summary, we have determined the upper critical field of single-crystal $NdFeAsO_{1-x}F_x$ using heat capacity measurements. The upper critical field slopes are $\partial H_{c2}^{c}/\partial T = -0.72$ T/K and $\partial H_{c2}^{ab}/\partial T = -3.1$ T/K, which correspond – in a single-band model – to zero-temperature coherence lengths of $\xi_{ab} \approx 3.7 nm$ and $\xi_c \approx 0.9 nm$ and a modest superconducting anisotropy of $\gamma \sim 4$. This anisotropy parameter is in good agreement with recent band structure calculations. In fields applied parallel to the c-axis the superconducting transition broadens significantly indicative of pronounced fluctuation effects. Therefore, we expect the appearance of an extended vortex liquid state and – in sufficiently clean samples – a vortex lattice melting transition, which are hallmarks of the phase diagram of the $CuO_2$-high-$T_c$ superconductors [23].

This work was supported by the US Department of Energy – Basic Energy Science – under contract DE-AC02-06CH11357, by the Natural Science Foundation of China, the Ministry of Science and Technology of China (973 project No. 2006CB60100, 2006CB921802, 2006CB921107) and the Chinese Academy of Sciences (Project ITSNEM).

Figure Captions

Fig. 1. Temperature dependence of the superconducting heat capacity, c(T,H)-c(T,7.5T||c), in various fields applied perpendicular and parallel to the FeAs-layers. The dashed lines indicate the shift of the peak and onset with increasing fields applied along the planes.

Fig. 2. Phase diagram of NdFeAsO$_{1-x}$F$_x$ as determined from field dependence of the peak positions and of T$_{c2}$. The dashed lines are linear fits yielding the average upper critical field slopes of -3.1 T/K and -0.72 T/K for the ab-plane and c-axis, respectively. Also included are the zero-resistance points in fields along the c-axis obtained on similar crystals [11].

Fig. 3. Superconducting heat capacity divided by temperature in 2 T || ab and 0.5 T || c. The dashed lines indicate the definitions of T$_p$ and T$_{c2}$. T$_{c2}$ is determined by the requirement that the triangular areas above and below the measured traces are equal. The inset shows the NdFeAsO$_{1-x}$F$_x$ crystallites mounted on the calorimeter. The Si$_3$N$_4$ membrane appears in blue, the Cu and Au/Co legs of the thermocouple are the vertical and horizontal metal films, and the diagonal lines are contacts to the meander heater located underneath the junction. The width of the thermocouple legs is about 100 μm.

Fig. 4. Angular dependence of the peak temperature in a field of 1.5 T. The red line is a fit with an anisotropy of γ = 4.

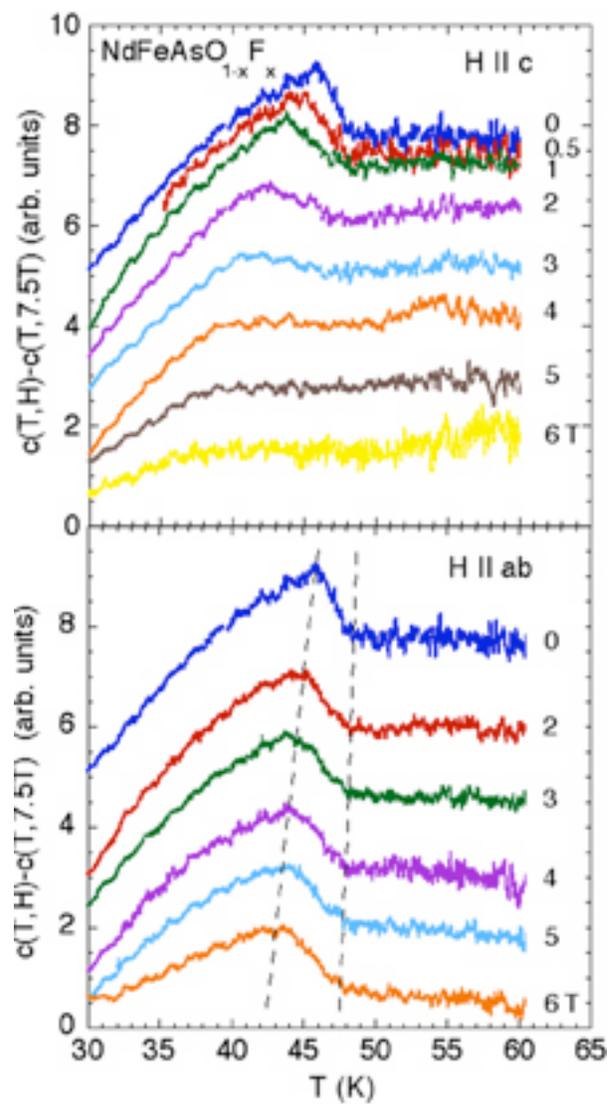

Figure 1

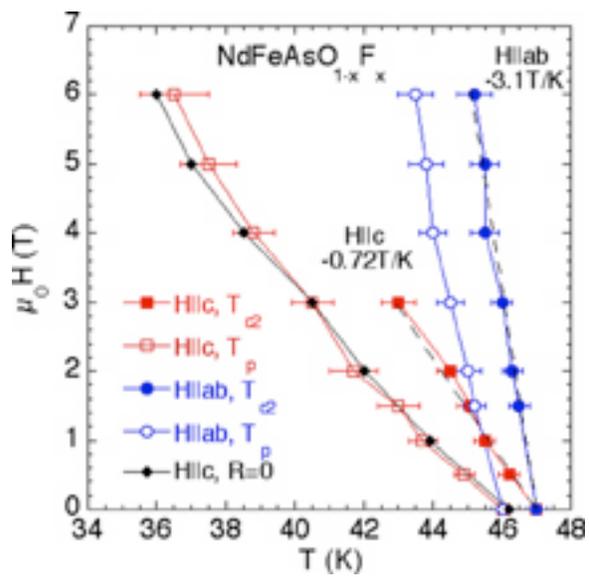

Figure 2

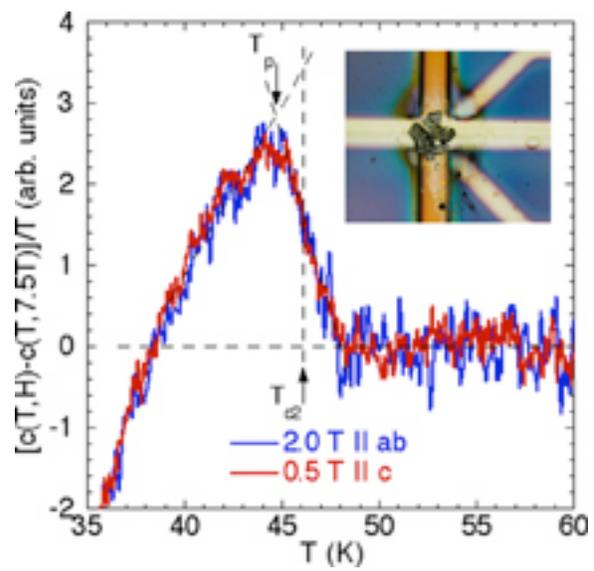

Figure 3

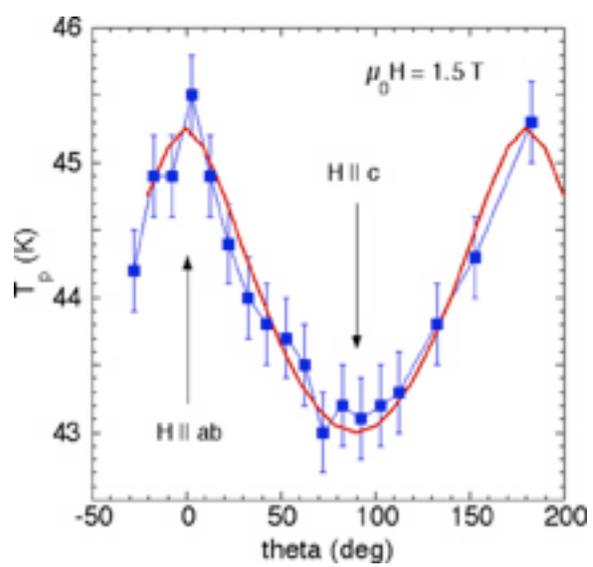

Figure 4